\documentclass[11pt]{article}

\usepackage[preprint]{acl}

\usepackage{times}
\usepackage{latexsym}
\usepackage{booktabs}
\usepackage{multirow}

\usepackage[T1]{fontenc}
\usepackage{bm}
\usepackage{xcolor}
\usepackage{listings}
\usepackage{tcolorbox}
\usepackage{graphicx}
\usepackage[table]{xcolor}
\definecolor{growthblue}{RGB}{239,246,251}
\definecolor{resultFail}{HTML}{B42318}
\definecolor{resultPass}{HTML}{237A47}
\tcbuselibrary{skins}

\definecolor{caseGray}{HTML}{596574}
\definecolor{caseBlue}{HTML}{38658A}
\definecolor{caseGreen}{HTML}{357361}

\lstdefinestyle{repaircase}{
  language=Python,
  basicstyle=\scriptsize\ttfamily,
  keywordstyle=\bfseries,
  stringstyle=\color{caseGray},
  showstringspaces=false,
  columns=fullflexible,
  keepspaces=true,
  breaklines=true,
  breakatwhitespace=false,
  frame=none,
  aboveskip=5pt,
  belowskip=0pt
}

\tcbset{
  repaircard/.style={
    enhanced,
    colback=white,
    boxrule=0.5pt,
    arc=1.5mm,
    left=5pt,
    right=5pt,
    top=5pt,
    bottom=5pt,
    toptitle=5pt,
    bottomtitle=5pt,
    fonttitle=\small\bfseries,
    coltitle=black,
    equal height group=repaircomparison
  }
}

\usepackage[utf8]{inputenc}
\usepackage{amsmath}
\usepackage{amssymb}
\usepackage{microtype}

\usepackage{inconsolata}

\usepackage{graphicx}

\title{SWE-growth: Teacher-Guided Adaptive Self-Play for Software Engineering Agents}

\author{
  Songshi Liang$^{1,2}$ \quad
  Zixuan Huang$^{3}$ \quad
  Mingxuan Song$^{4}$ \quad
  Haoyu Tang$^{1}$ \\
  \textbf{Shusheng Xu}$^{2}$ \quad
  \textbf{Tongkai Yang}$^{2}$ \quad
  \textbf{Zihe Wang}$^{1}$\thanks{
    Corresponding authors:
    Zihe Wang (\texttt{wang.zihe@ruc.edu.cn}) and
    Rui Yan (\texttt{rui.yan@whu.edu.cn}).
  } \quad
  \textbf{Rui Yan}$^{5,6}$\footnotemark[1]
  \\[0.4em]
  \small
  $^{1}$Gaoling School of Artificial Intelligence,
  Renmin University of China \\
  \small
  $^{2}$Ant Group \quad
  $^{3}$Beihang University \quad
  $^{4}$Peking University \\
  \small
  $^{5}$School of Artificial Intelligence,
  Wuhan University \\
  \small
  $^{6}$School of Computer Science,
  Wuhan University
}

\begin{document}
\maketitle
\begin{abstract}
Self-play offers a scalable way to train software engineering
agents by generating repository-level repair tasks.
However, with binary repair rewards and finite
rollout budgets, all-fail groups provide no
reward-based group-relative advantage. Such tasks may still offer learning opportunities
when a Teacher can provide verified repairs
absent from the Student's own attempts.
We introduce \textsc{SWE-Growth}, a teacher-guided adaptive
self-play framework that uses verified Teacher outcomes
to identify learning opportunities for the current Student.
Tasks with all students failure but reliable Teacher
success provide demonstrations for supervised fine-tuning (SFT), and post-SFT Student rollouts identify
mixed-outcome tasks with a positive Teacher--Student gap
for reinforcement learning.
Task eligibility also guides Injector updates, connecting
Student training with subsequent task generation.
Starting from 101 seed issues, \textsc{SWE-Growth} improves
Qwen3-Coder-30B-A3B-Instruct from 27.7\% to 31.7\%
on SWE-bench Pro and from 11.0\% to 17.2\%
on held-out TerminalBench 2.1 by Round~2.
It exceeds Student-only self-play by 2.6 and 4.1 percentage
points, respectively, and Teacher-augmented self-play
by 1.8 and 2.5 points under approximately matched budgets.
These results support adapting both supervision and task
selection to the Student's evolving repair capability.

\end{abstract}

\section{Introduction}

Large language model (LLM)-based software engineering agents
solve repository-level issues by exploring codebases, editing
source code, and checking their changes through execution
\citep{DBLP:conf/iclr/JimenezYWYPPN24,NEURIPS2024_5a7c9475}.
These test results provide feedback for training agents to
produce better repairs.
However, training typically relies on prepared tasks,
executable repository environments, and successful repair
examples~\citep{NEURIPS2025_8b86cf5a,jain2025r2egymproceduralenvironmentshybrid}.
As an agent improves, some tasks become too easy, while others
remain beyond its current ability.
This motivates generating and selecting training tasks that
adapt to what the agent can learn next.

Self-play offers a way to make training experience coevolve with the agent~\citep{NEURIPS2025_9837dc00}.
\textbf{Yet generating new challenges does not necessarily yield new learning opportunities.} 
Under binary repair rewards, tasks on which all
sampled attempts succeed or fail provide zero
group-relative advantages~\citep{shao2024deepseekmath}.
Prior work addresses related learning bottlenecks
by combining external demonstrations with reinforcement
learning, interleaving fine-tuning on difficult tasks,
or deriving denser supervision from feedback
~\citep{yan2026learning,ma2026learning,hubotter2026reinforcement}. Different from these works, we explicitly use
task-level gaps between a frozen Teacher and
the current Student to allocate supervision.
This design brings advantage that supervision targets tasks
where the Teacher demonstrates a verified local
strength that the Student has yet to acquire.
As the Student improves, these gaps are reassessed
to keep guidance focused on its remaining needs.

Consequently, we propose \textsc{SWE-Growth}, a
repository-level self-play framework that uses
task-level Teacher--current Student gaps to coordinate
SFT, GRPO, and task generation.
\textbf{The Teacher supplies verified repair
demonstrations and helps identify tasks on which
the current Student has room to learn.}
An Injector generates execution-validated bugs,
which the Student and frozen Teacher attempt to repair.
Tasks with all-fail Student groups and reliable
Teacher successes provide demonstrations for SFT.
After SFT, fresh Student rollouts identify
mixed-outcome tasks with a positive
Teacher--Student gap for GRPO.
Curriculum eligibility also rewards the Injector,
connecting Student improvement with subsequent
bug generation while the Teacher remains frozen.
This coupling relates to the broader study of
recursive self-improvement \citep{tan2026metarsi}.

We evaluate \textsc{SWE-Growth} on SWE-bench Pro
\citep{deng2025swe} and assess cross-domain generalization
on held-out TerminalBench 2.1 \citep{merrill2026terminal}.
Starting from 101 seed issues, the method improves the
Student backbone from 27.7\% to 31.7\% on SWE-bench Pro
and from 11.0\% to 17.2\% on TerminalBench 2.1 by Round~2.
These scores exceed our same-backbone Student-only
self-play baseline by 2.6 and 4.1 percentage points,
respectively.
To distinguish the value of adaptive curriculum design
from access to Teacher demonstrations, we also compare
against Teacher-augmented self-play under approximately
matched resource budgets.
\textsc{SWE-Growth} retains gains of 1.8 and 2.5 percentage
points in this comparison.
We further investigate task-level complementarity using
a weak Teacher
and use ablations to examine the roles of SFT, GRPO,
and refreshed task selection.
Together, these experiments assess both the effectiveness
of the framework and the contribution of its curriculum
design.

\begin{table*}[t]
\centering
\caption{
Teacher supervision exposes learning opportunities within silent
Student self-play groups.
}
\label{tab:motivation_signal}
\small
\setlength{\tabcolsep}{3pt}
\renewcommand{\arraystretch}{1.08}

\begin{tabular*}{\textwidth}{
  @{\extracolsep{\fill}}
  ccccccc
  @{}
}
\toprule

\multicolumn{4}{c}{
  \textbf{Student-only GRPO} ($N=63$)
}
&
\multicolumn{2}{c}{
  \textbf{Teacher on all-fail} ($N=47$)
}
&
\multicolumn{1}{c}{
  \textbf{Selected subset}
}
\\

\cmidrule(lr){1-4}
\cmidrule(lr){5-6}
\cmidrule(l){7-7}

All-fail
&
All-pass
&
Mixed
&
Silent
&
$T\!\geq\!1/4$
&
$T\!\geq\!2/4$
&
$0/4 \rightarrow$ mixed
\\

\midrule

47 (74.6\%)
&
3 (4.8\%)
&
13 (20.6\%)
&
\textbf{50 (79.4\%)}
&
17 (36.2\%)
&
\textbf{9 (19.1\%)}
&
\textbf{3/5 (60.0\%)}
\\

\bottomrule
\end{tabular*}

\vspace{2pt}
\parbox{\textwidth}{
\footnotesize
\textit{Notes.}
A group is \emph{silent} when its four Student rollouts receive
identical rewards and therefore have zero group-relative advantage.
$T\!\geq\!k/4$ means that the Teacher succeeds in at least $k$ of four
attempts.
}
\end{table*}
\begin{figure*}
    
    \centering

    \begin{minipage}[c]{0.55\textwidth}
        \centering
        \includegraphics[width=\textwidth]
        {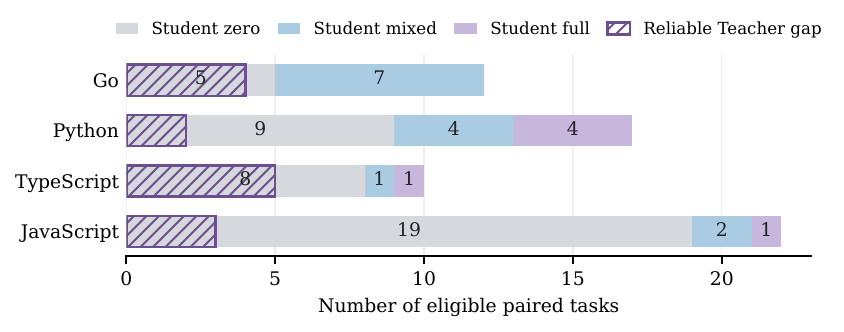}
    \end{minipage}
    \hspace{0.045\textwidth}
    \begin{minipage}[c]{0.36\textwidth}
        \centering
        \includegraphics[width=\textwidth]
        {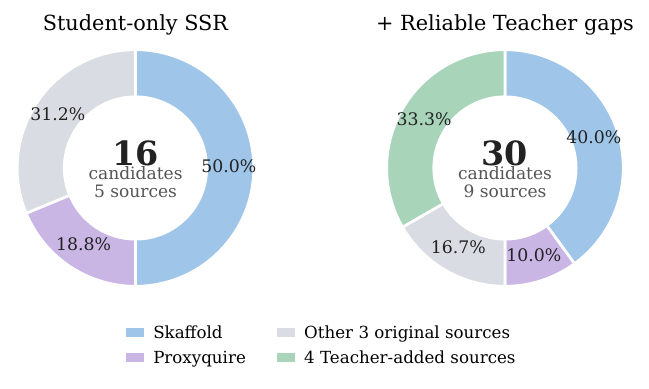}
    \end{minipage}

    \caption{
    Student-only self-play leaves sparse and concentrated learning
    signals, while a frozen Teacher expands the reachable signal.
    \textbf{Left:} Student outcomes by language on complete,
     groups; hatching marks reliable Teacher gaps
    within the Student-zero region.
    \textbf{Right:} source composition before and after adding reliable
    Teacher gaps.
    }
    \label{fig:ssr_teacher_curriculum}
    \vspace{-0.6\baselineskip}
\end{figure*}
\section{Related work}
SWE-Gym and R2E-Gym provide executable environments for training
software agents and verifiers
~\citep{pmlr-v267-pan25g,jain2025r2egymproceduralenvironmentshybrid}.
SWE-smith scales synthetic repository bug generation
~\citep{NEURIPS2025_8b86cf5a}, while BugPilot generates bugs through
feature-development workflows to better approximate naturally
occurring errors~\citep{sonwane2025bugpilotcomplexbuggeneration}.
Beyond expanding task supply, Self-Play SWE-RL jointly trains
bug-injection and repair roles, using the current Solver's repair
rate to guide the injection curriculum
~\citep{wei2026trainingsuperintelligentsoftwareagents}.
Socratic-SWE extracts skills from solving traces and prioritizes
tasks by gradient alignment with a trusted validation set
~\citep{xiao2026socraticsweselfevolvingcodingagents}, while Anchored
Self-Play mitigates synthetic-distribution drift through
reference-guided generation and reference-example replay
~\citep{choi2026anchoredselfplaycoderepair}. 
However, generating valid repair tasks does not
ensure that they provide useful learning signals
for the current Student.
LUFFY and ReLIFT address related bottlenecks by
combining external demonstrations with reinforcement
learning~\citep{yan2026learning,ma2026learning},
while AgentFrontier uses capability differences
to select challenging training
tasks~\citep{chen2025agentfrontier}.
Building on these ideas, \textsc{SWE-Growth}
uses task-level Teacher--current Student gaps
to select SFT demonstrations, reassesses Student
outcomes after SFT to select mixed groups with
a positive gap for GRPO, and feeds curriculum
eligibility back into Injector updates.
Appendix~\ref{sec:add_related_work}
provides a more detailed comparison with adaptive
curricula, capability-calibrated task generation,
Teacher-guided learning and self-distillation.

\section{Motivation: From Task Generation to Learning Signals}
\label{sec:motivation}

Self-play can generate new repair tasks, but can the
Student learn from them?
Drawing on findings from LUFFY and ReLIFT that
external demonstrations can complement learning
from a model's own rollouts in RL
~\citep{yan2026learning,ma2026learning},
we further investigate this issue in repository-level
self-play through task-level diagnostics.
We use eight source tasks specifically per language from Go, Python,
TypeScript, and JavaScript.
With $G=4$ repair attempts per injected task, we analyze
63 complete Student groups across two rollout seeds,
evaluate a frozen Teacher under matched per-attempt budgets,
and reassess selected tasks after demostration.

\paragraph{Most generated tasks provide no group-relative
learning signal.}
Only 13 of 63 Student groups (20.6\%) contain both
successful and unsuccessful repairs.
The remaining 47 all-fail and 3 all-pass groups
produce identical within-group rewards and therefore
zero group-relative advantages
(Table~\ref{tab:motivation_signal}).
Thus, generating tasks does not ensure useful GRPO
comparisons: 79.4\% of the observed groups provide
no reward-advantage signal.
The usable groups are also concentrated in a few
source issues despite language-balanced sampling
(Figure~\ref{fig:ssr_teacher_curriculum}).
The bottleneck is therefore not simply obtaining
more tasks, but obtaining repair attempts from
which the Fixer can learn.

\paragraph{Teacher demonstrations can make failed tasks
useful for subsequent reinforcement learning.}
Are all-fail tasks beyond reach, or does the Student
lack successful examples to learn from?
The frozen Teacher succeeds at least once in four
attempts on 17 of the 47 tasks (36.2\%), including
9 with at least two successes.
These results identify verified repair examples
where Student-only sampling produces none.
The complementary source-coverage analysis also
shows that Teacher guidance reaches source issues
absent from the Student's mixed-outcome pool
(Figure~\ref{fig:ssr_teacher_curriculum}). After 12 SFT steps on Teacher demonstrations,
three of five previously all-fail tasks in the
Teacher-selected subset evaluated before and after
SFT yield mixed Student outcomes.
On these tasks, the Student can now generate its own
successful and unsuccessful repairs, providing the
comparisons needed for GRPO.
Student successes also increase from 195 to 204
on 722 paired SWE-bench Pro issues, suggesting
transfer beyond the injected tasks.
These observations support using task-level
Teacher--current Student gaps to guide curriculum
selection in repository self-play.
Verified Teacher successes identify demonstrations
for selected Student-zero tasks; after SFT,
reassessment identifies mixed-outcome tasks with
a remaining positive gap for GRPO.

\begin{figure*}[t]  
    \centering
    \includegraphics[width=0.92\linewidth]{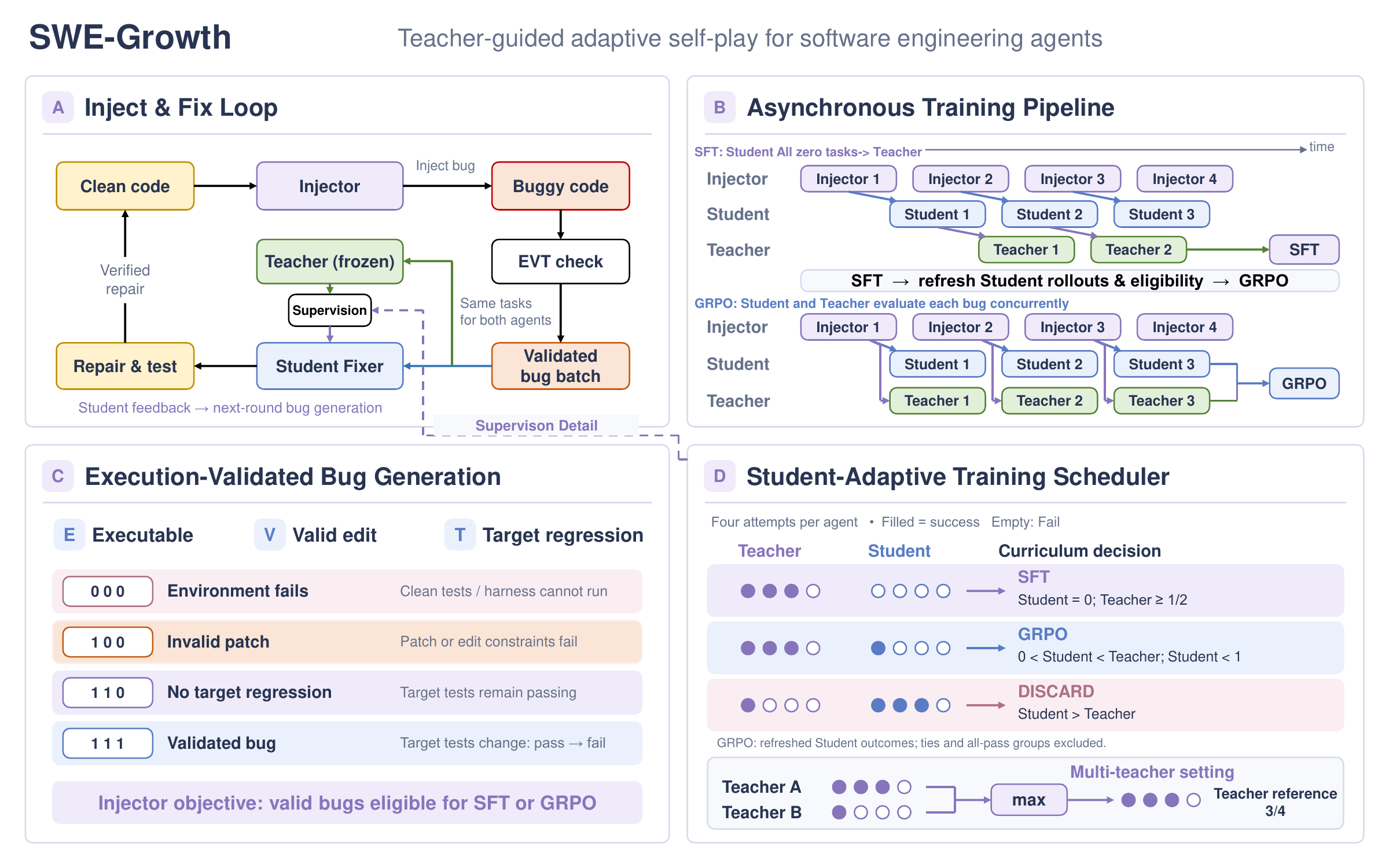}  
    \caption{Main Framework of our SWE-Growth}
    \label{SWE_Growth_Main}
\end{figure*}
\section{Methods}

\subsection{Problem Formulation}

Let $t \in \mathcal{T}$ denote a clean repository task,
consisting of a repository revision, task context, and designated
tests.
An Injector policy $\pi_{\mathrm{inj}}$ proposes a bug patch $b$,
producing a synthetic repair task $t_b$.
A Student Fixer $\pi_{\mathrm{stu}}$ interacts with the modified
repository through tools and submits a repair patch $p$.
A frozen Teacher Fixer $\pi_{\mathrm{tea}}$ attempts the same task
under a matched interaction budget.
The Injector and Student Fixer are initialized from the same base model
and subsequently updated using their role-specific
rewards.
The objective is to construct a training curriculum that improves
the Student's repair performance on held-out real issues.

\subsection{Learning to inject repository bugs.}
Drawing on repository bug-generation approaches
in SWE-smith and BugPilot
~\citep{NEURIPS2025_8b86cf5a,sonwane2025bugpilotcomplexbuggeneration},
we construct injection examples from a multilingual
collection of resolved issues with executable
environments, test specifications, and developer patches.
At each round, we collect trajectories associated
with tasks the Student failed to solve in the
preceding round and construct new bugs using
reverse-PR and related transformations.
For initialization, this process uses trajectories
from the initial Student and provides supervised
examples for training the Injector before self-play
generation begins.

For each source task $t$, we restore the repository
to its clean, repaired state and provide the trained
Injector with the repository context and issue
specification.
The Injector then generates an injection patch
\begin{equation}
b \sim \pi_{\mathrm{inj}}(\cdot \mid t),
\end{equation}
which modifies production code to create a
synthetic repair task $t_b$.
Only injections passing the execution checks below
enter the repair curriculum.

\subsection{EVT validation of injected bugs.}
We validate each proposed injection through three sequential checks.
Executability $E(t)$ requires the clean repository and its test harness
to run successfully, with the target tests initially passing.
Conditional on $E(t)=1$, validity $V(t,b)$ requires the injection patch
to apply and satisfy the prescribed edit constraints.
Conditional on $E(t)=V(t,b)=1$, target failure $T(t,b)$ requires the
injected repository to change the target tests from passing to failing:
\begin{equation}
Q(t,b)=E(t)\,V(t,b)\,T(t,b).
\end{equation}
A task is retained only when $Q(t,b)=1$, the designated non-target
tests remain passing, and the patch satisfies the non-catastrophic-edit
constraints.
These checks retain injections that cause a
target-test regression while preserving the
designated non-target tests.
Repair outcomes then guide curriculum selection.

\subsection{Teacher--Student Evaluation and Learning}

SWE-Growth combines Teacher demonstrations with Student reinforcement
learning to turn validated injections into an adaptive curriculum.
The Teacher remains frozen, while repair outcomes guide both
Student learning and Injector updates.

\paragraph{Repair evaluation and curriculum routing.}
For each validated task $t_b$, the Student performs $G$ independent
repair rollouts. ($G$=4 by default)
A candidate patch receives reward:
\begin{equation}
r(t_b,p)=A(t_b,p)\,U(t_b,p)\,P(t_b,p),
\end{equation}
where the binary indicators denote successful patch application,
target-test restoration, and non-target-test preservation. A candidate receives positive reward only when all three conditions meet.
For agent $a$ at round $k$, we estimate:
\begin{equation}
\widehat{s}_{a,k}(t_b)
=
\frac{1}{G}\sum_{g=1}^{G}r(t_b,p_{a,g}).
\end{equation}
Tasks with Student success rates below one are evaluated by
$M$ frozen Teachers, each using $G$ rollouts under the same
task input, tools, and interaction budget.
The single-Teacher setting corresponds to $M=1$.
Using complete groups with valid execution results, we define the teacher-student capability gap as:
\begin{equation}
\begin{aligned}
\widehat{s}_{\mathrm{ref},k}(t_b)
&= \max_{1 \leq m \leq M}
   \widehat{s}_{\mathrm{tea}_m,k}(t_b), \\
\widehat{\Delta}_k(t_b)
&= \widehat{s}_{\mathrm{ref},k}(t_b)
 - \widehat{s}_{\mathrm{stu},k}(t_b).
\end{aligned}
\end{equation}
We first construct the SFT pool from Student-zero tasks
with reliable Teacher successes:
\begin{equation}
\mathcal{D}_{\mathrm{SFT},k}
=
\left\{
t_b:
\widehat{s}^{\mathrm{pre}}_{\mathrm{stu},k}(t_b)=0,\;
\widehat{s}_{\mathrm{ref},k}(t_b)\geq\frac{1}{2}
\right\}.
\end{equation}
With $G=4$, this requires at least one Teacher to succeed
in two of its four attempts.
We extract verified successful demonstrations from Teachers
meeting this threshold.

The GRPO pool contains tasks with mixed Student outcomes
and a positive Teacher--Student gap:
\begin{equation}
\begin{aligned}
\widehat{\Delta}^{\mathrm{post}}_k(t_b)
&=
\widehat{s}_{\mathrm{ref},k}(t_b)
-
\widehat{s}^{\mathrm{post}}_{\mathrm{stu},k}(t_b),\\
\mathcal{D}_{\mathrm{GRPO},k}
&=
\left\{t_b:
\begin{aligned}
&0<\widehat{s}^{\mathrm{post}}_{\mathrm{stu},k}(t_b)<1,\\
&\widehat{\Delta}^{\mathrm{post}}_k(t_b)>0
\end{aligned}
\right\}.
\end{aligned}
\end{equation}
The superscripts pre and post denote Student
estimates before and after SFT, respectively.
These tasks provide both successful and unsuccessful Student
trajectories, with at least one Teacher achieving a higher
success rate.

\paragraph{Demonstration learning and reinforcement.}
Our main configuration uses action-level decision records
for both SFT and GRPO. SFT optimizes retained Teacher actions conditioned
on their preceding interaction histories.
Only the action tokens are optimized, while the history
provides conditioning context.
Appendix~\ref{Training details} details the record construction and loss masking.
Teacher demonstrations and successful replay examples
define the SFT objective:
\begin{equation}
\mathcal{D}_{k}^{\mathrm{mix}}
=
(1-\lambda_{\mathrm{rep}})\mathcal{D}_{\mathrm{demo},k}
+
\lambda_{\mathrm{rep}}\mathcal{D}_{\mathrm{replay},k}.
\end{equation}
\begin{equation}
\mathcal{L}_{\mathrm{SFT}}(\theta)
=
-\underset{(h,a)\sim\mathcal{D}_{k}^{\mathrm{mix}}}
{\mathrm{E}}\,
\log \pi_\theta(a\mid h).
\end{equation}

After SFT, the updated Student attempt the selected tasks
again. A task enters GRPO if these new attempts include both
successes and failures and its success rate remains below the
frozen Teacher's estimate. Appendix~\ref{app:grpo_details} defines the advantage
estimator, policy ratio, KL reference and penalty,
loss masking, and retention-anchor objective. Let $\mathcal{V}_g$ denote the valid action-token positions
across all retained decision records of rollout $g$,
where $d$ indexes records and $j$ indexes action tokens.
Each record shares the advantage $\widehat{A}_g$
of its originating rollout. We define
\begin{equation}
\begin{aligned}
\phi(\rho,A)
&=
\min\{\rho A,\operatorname{clip}
(\rho,1-\epsilon,1+\epsilon)A\},\\
\ell_{g,d,j}(\theta)
&=
\phi\!\left(\rho_{g,d,j}(\theta),\widehat{A}_g\right)
-\beta K_{g,d,j}(\theta).
\end{aligned}
\end{equation}
We average over valid action tokens within each rollout
and then equally across the $G$ rollouts: 
\begin{equation}
\mathcal{L}_{\mathrm{GRPO}}(\theta)
=
-\mathbb{E}_{\mathcal{G}}
\left[
\frac{1}{G}
\sum_{g=1}^{G}
\frac{1}{|\mathcal{V}_g|}
\sum_{(d,j)\in\mathcal{V}_g}
\ell_{g,d,j}(\theta)
\right].
\end{equation}
In Appendix~\ref{sec:rollout_group_size}, we also discuss the performance and group composition when $G=8$.
\paragraph{Injector feedback.}
A valid injection earns reward if its task enters the SFT or
GRPO pool; an invalid injection is penalized:
\begin{equation}
\begin{aligned}
R_{\mathrm{inj},k}(t,b)
&=Q(t,b)F_k(t_b)\\
&\quad-(1-Q(t,b))c_{\mathrm{EVT}}(t,b)
\end{aligned}
\end{equation}
Here, $Q$ indicates a valid injection and $F_k$ indicates
SFT or GRPO eligibility. The Fixer receives the repair reward.
The penalty $c_{\mathrm{EVT}}$ is $1.0$ for non-executable or
catastrophic injections, $0.4$ for evaluable but invalid patches,
and zero for otherwise valid edits without a target regression.
Selection outcomes are fixed for each update and refreshed as
the Student changes. 

\subsection{Two-Stage Asynchronous Execution}

We implement an asynchronous schedule in which
the Injector generates bugs while the Student
and Teacher evaluate previously generated tasks.
Each validated injection enters the Student queue;
complete Student-zero and mixed-outcome groups then
enter the Teacher queue while evaluation of other
tasks continues.
Teacher results determine task eligibility for
demonstration learning and reinforcement learning.
Fig~\ref{SWE_Growth_Main} illustrates the asynchronous process and
Appendix~\ref{async} reports the execution-efficiency comparison.

\begin{table*}[t]
\centering
\small
\caption{
Performance on SWE-bench Pro Public (731 tasks) and
TerminalBench 2.1, reported as mean $\pm$ standard
deviation across three evaluation seeds.
$^{*}$ denotes significant improvement over Round-2
Self-Play SWE-RL + Teacher
(paired $t$-test across evaluation seeds, $p<0.05$).
Training and evaluation details appear in Appendix~B.
}
\setlength{\tabcolsep}{7pt}
\renewcommand{\arraystretch}{1.05}

\begin{tabular}{@{}lccc@{}}
\toprule
Model / Method & Round
& \shortstack{SWE-bench Pro\\Resolve rate (\%) $\uparrow$}
& \shortstack{TerminalBench 2.1\\Success rate (\%) $\uparrow$} \\
\midrule

\multicolumn{4}{@{}l}{\textit{Backbone and baselines without self-play}} \\
\addlinespace[2pt]
Qwen3-Coder-30B-A3B-Instruct & --
& $27.7 \pm 1.3$ & $11.0 \pm 1.7$ \\
Qwen3.6-35B-A3B (Teacher reference) & --
& $38.5 \pm 1.9$ & $42.1 \pm 2.4$ \\
SWE-MeM SFT+RL & --
& $31.1 \pm 1.1$ & -- \\
Teacher Distillation$^{\dagger}$ & --
& $30.8 \pm 0.9$ & $9.8 \pm 2.2$ \\

\midrule
\multicolumn{4}{@{}l}{\textit{Student-only self-play}} \\
\addlinespace[2pt]
Self-Play SWE-RL$^{\dagger}$ & 0
& $28.2 \pm 1.1$ & $11.8 \pm 1.7$ \\
& 1 & $28.6 \pm 1.3$ & $12.7 \pm 1.6$ \\
& 2 & $29.1 \pm 1.3$ & $13.1 \pm 2.1$ \\

\midrule
\multicolumn{4}{@{}l}{\textit{Teacher-guided self-play}} \\
\addlinespace[2pt]
Self-Play SWE-RL + Teacher$^{\dagger}$ & 0
& $28.4 \pm 1.4$ & $13.1 \pm 1.7$ \\
& 1 & $29.2 \pm 1.1$ & $14.0 \pm 1.3$ \\
& 2 & $29.9 \pm 1.4$ & $14.7 \pm 1.9$ \\

\addlinespace[3pt]
\rowcolor{growthblue}
\textsc{SWE-Growth} (Ours) & 0
& $28.5 \pm 1.3$ & $13.8 \pm 2.2$ \\
\rowcolor{growthblue}
& 1
& $\mathbf{30.2 \pm 1.1}$
& $\mathbf{16.0 \pm 1.9}$ \\
\rowcolor{growthblue}
& 2
& $\mathbf{31.7 \pm 1.1}^{*}$
& $\mathbf{17.2 \pm 2.2}^{*}$ \\
\bottomrule
\end{tabular}

\begin{minipage}{0.90\textwidth}
\footnotesize
\textit{Notes.}
$^{\dagger}$ Our implementation.
In teacher distillation setting, we distill from over 6500 pieces of instance from SWE-smith. 
\label{main_table}
\end{minipage}
\end{table*}
\section{Results}

\subsection{Main Results and Findings}

We evaluate repository-level repair on SWE-bench Pro
and cross-domain generalization on TerminalBench 2.1,
without training on TerminalBench tasks.
All controlled training comparisons use
Qwen3-Coder-30B-A3B-Instruct~\citep{cao2026qwen3} as the Student backbone.
We compare SWE-Growth with training without self-play
(SWE-MeM SFT+RL~\citep{gao2026swe} and Teacher Distillation),
Student-only self-play (Self-Play SWE-RL),
and Teacher-guided self-play.
Qwen3.6-35B-A3B~\citep{team2026qwen3} serves as the default frozen Teacher. We conduct experiments of more teachers in Appendix~\ref{more_teachers}. The details of more SWE-bench Pro and TerminalBench 2.1 harness are in Appendix~\ref{evaluation_details}.  

Self-Play SWE-RL + Teacher adds verified Teacher
demonstrations for SFT on all student failed tasks and then apply GRPO to all Student mixed-outcome tasks,
without requiring a positive Teacher--Student gap,
and retains the original self-play Injector reward.
This baseline and \textsc{SWE-Growth} start from
the same Student and Injector checkpoints and
continue from their own checkpoints across rounds
under approximately matched resource budgets as shown in Table~\ref{tab:teacher_budget_usage} in the Appendix. Round 0 denotes for distillation from Kimi-k3's~\citep{team2026kimi} rollout from the original student backbone.
Further training configurations
and resource usage are provided in Appendix~\ref{Training details}. In Appendix~\ref{sec:additional_ablation} we also discuss the results when we discard choosing GRPO groups with positive gap and warmup from Kimi-k3's rollout.

\noindent\textbf{SWE-Growth achieves sustained improvements.}
From Tab~\ref{main_table}, we find that starting from 101 source tasks, SWE-Growth improves
SWE-bench Pro performance from the base model's
27.7\% to 28.5\%, 30.2\%, and 31.7\% across
Rounds~0--2.
TerminalBench performance also increases from
11.0\% to 13.8\%, 16.0\%, and 17.2\%.
SWE-Growth exceeds Student-only self-play
by 2.6 and 4.1 percentage points on the two
benchmarks, respectively.

\noindent\textbf{Adding a Teacher does not explain the full gain.}
Self-Play SWE-RL + Teacher improves over Student-only
self-play, reaching 29.9\% on SWE-bench Pro and
14.7\% on TerminalBench in Round~2.
However, SWE-Growth maintains an advantage that
widens across rounds: from 0.1 to 1.8 percentage
points on SWE-bench Pro, and from 0.7 to 2.5 points
on TerminalBench.
This comparison supports the combined value of
SWE-Growth's curriculum selection and Injector
feedback beyond adding Teacher demonstrations
to the baseline.

\noindent\textbf{Repair gains extend beyond demonstration learning.}
Teacher Distillation reaches 30.8\% on SWE-bench Pro,
but its TerminalBench score falls from the base
model's 11.0\% to 9.8\%.
Thus, demonstration learning alone
does not support the cross-domain gains.
SWE-Growth also achieves a higher SWE-bench Pro
point estimate than SWE-MeM SFT+RL
(31.7\% versus 31.1\%), despite starting from
101 source tasks rather than thousands of source
training problems.
Together, these results demonstrate the potential
of Teacher-guided self-play to support continued
improvement from a small initial task collection.

There remains a question of where \textsc{SWE-Growth}'s
improvement comes from.
\emph{Does the Student merely reproduce the Teacher's successes,
or does teacher-guided self-play also expand its coverage
to tasks neither reference model solves?
We conduct further experiments in Sec ~\ref{Weak Teacher}.}

\begin{figure*}[t]
    \centering
    \includegraphics[width=0.95\textwidth]
    {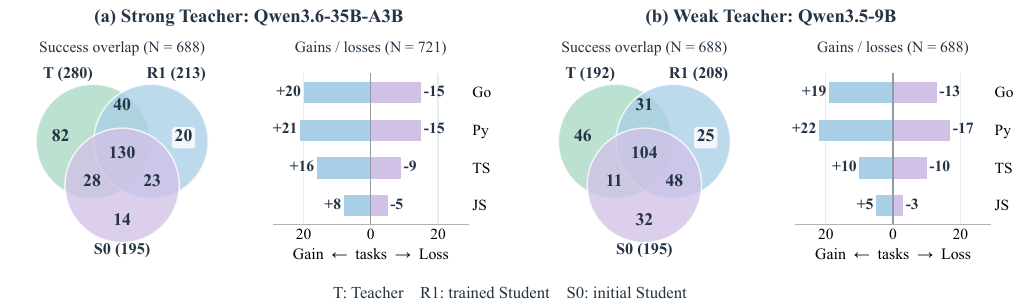}
    \caption{
    Repair coverage with (a) a strong Teacher and
    (b) a weak Teacher.
    Each panel shows success overlap among the Teacher
    (T), trained Student (R1), and initial Student (S0),
    followed by language-wise gains and losses.
    Highlighted counts denote tasks solved only by R1.
    Each overlap diagram uses 688 tasks completed by
    all three models; language comparisons use 721
    paired tasks in (a) and 688 in (b).
    Circle areas are schematic.
    }
    \label{round1_gains}
\end{figure*}
\subsection{Learning from the Teacher, growing beyond its coverage.}
\label{Weak Teacher}
We examine whether SWE-Growth expands repair coverage
beyond the Teacher's observed successes and whether
useful guidance requires higher aggregate Teacher
performance.
Figure~\ref{round1_gains} compares the initial Student
(S0), trained Student (R1), and Teacher (T).
Language-wise gains and losses compare the two Student
checkpoints on 721 commonly completed tasks in the
strong-Teacher setting and 688 in the weak-Teacher setting. In both settings, \textbf{we train the student directly from the backbone Qwen3-Coder-30B-A3B-Instruct without Kimi-k3 warmup}.

With the strong Qwen3.6-35B-A3B Teacher, the Round-1
Student solves 20 tasks that neither the initial
Student nor the Teacher solves on the 688 tasks
commonly evaluated by all three models
(Figure~\ref{round1_gains}(a)).
The language-wise comparison instead uses the larger
721-task subset evaluated by both Student checkpoints:
it gains 65 successes while losing 44, with gains
exceeding losses in all four languages.
Thus, the improvement extends beyond the Teacher's
observed success set and is not confined to one language.

We further examine whether useful guidance requires
a Teacher with higher aggregate performance.
Qwen3.5-9B solves 192 of 731 SWE-bench Pro tasks,
compared with 199 for the initial Student.
A closer comparison of paired training rollouts
also finds fewer Teacher-only successes than
Student-only successes.
Together this
motivates using Qwen3.5-9B as our weak Teacher.
We apply the same SFT--GRPO pipeline using verified
Teacher gaps, and find that
on the 688-task common subset
in Figure~\ref{round1_gains}(b), the trained Student
solves 208 tasks, exceeding the initial Student's
195 and the Teacher's 192. 
Additionally, the trained Student solves 25 tasks missed by both
original models, spanning four languages and ten
repositories.
Together, these findings suggest that a Teacher's
task-level strengths can support Student improvement
even without aggregate superiority, and that the
resulting gains can extend beyond both models'
initial observed repair coverage.
\begin{table*}
\centering
\caption{
Ablation results by training round.
In each round, all training variants start from the same best
checkpoint selected in the preceding round and differ only in
the current round's training strategy.
}
\label{tab:ablation}
\small
\setlength{\tabcolsep}{5pt}
\renewcommand{\arraystretch}{1.1}
\resizebox{0.85\textwidth}{!}{%
\begin{tabular}{llcccc}
\toprule
Round & Method
& SWE-bench Pro (\%) $\uparrow$
& Terminal-Bench (\%) $\uparrow$
& CCR (\%) $\uparrow$
& CMR (\%) $\downarrow$ \\
\midrule

\multirow{6}{*}{Round 1}
& Base Model
& $27.7 \pm 1.3$ & $11.0 \pm 1.7$ & --- & --- \\
\midrule
& \textbf{SWE-Growth}
& $\mathbf{30.2 \pm 1.1}$ & $\mathbf{16.0 \pm 1.9}$
& 100.0 & 0.0 \\
& vs. SFT only
& $29.1 \pm 0.9$ & $11.3 \pm 1.1$ & --- & --- \\
& vs. GRPO only
& $28.5 \pm 1.3$ & $12.0 \pm 1.3$ & 100.0 & 0.0 \\
& vs. No refresh
& $28.7 \pm 1.3$ & $11.3 \pm 0.9$ & 58.1 & 15.1 \\
& vs. Fixed-pair gap
& $28.2 \pm 0.8$ & $11.0 \pm 1.1$ & 29.4 & 42.4 \\
\midrule

\multirow{5}{*}{Round 2}
& \textbf{SWE-Growth}
& $\mathbf{31.7 \pm 1.1}$ & $\mathbf{17.2 \pm 2.2}$
& 100.0 & 0.0 \\
& vs. SFT only
& $30.4 \pm 1.6$ & $13.7 \pm 1.9$ & --- & --- \\
& vs. GRPO only
& $31.4 \pm 1.3$ & $16.5 \pm 1.5$ & 100.0 & 0.0 \\
& vs. No refresh
& $31.0 \pm 1.1$ & $16.0 \pm 1.3$ & 78.5 & 9.5 \\
& vs. Fixed-pair gap
& $30.8 \pm 1.6$ & $15.4 \pm 2.1$ & 44.3 & 44.9 \\
\bottomrule
\end{tabular}%
}
\end{table*}

\subsection{Ablations of Student Training and Teacher Guidance}

We compare training strategies through single-round
ablations from shared Student checkpoints.
Round-1 variants share the checkpoint used to start
standard SWE-Growth Round-1 training.
Round-2 variants all start from the standard
SWE-Growth Round-1 checkpoint.
These experiments measure each strategy's effect
at a common starting capability within each round,
not its cumulative effect across rounds.

We conduct four ablations: \textbf{SFT only} omits GRPO, while
\textbf{GRPO only} omits SFT.
\textbf{No refresh} retains pre-SFT task selections
but collects new post-SFT Student rollouts.
\textbf{Fixed-pair gap} uses the gap between frozen
Qwen3.6 and Qwen3.5 models instead of the
Teacher--current Student gap, while retaining
the current-Student mixed-outcome requirement. To better assess curriculum selection, we design two additional metrics. Let $\mathcal{C}_k$ denote the selected GRPO pool
and $\mathcal{E}_k$ the tasks in the common candidate
pool satisfying
\begin{equation}
\begin{aligned}
&0<\widehat{s}^{\mathrm{cur}}_{\mathrm{stu},k}(t)<1,\\
&\widehat{s}_{\mathrm{ref},k}(t)
>\widehat{s}^{\mathrm{cur}}_{\mathrm{stu},k}(t).
\end{aligned}
\end{equation}
We assess eligibility using each variant's Student
immediately before GRPO: the starting checkpoint
for GRPO only and the post-SFT checkpoint otherwise.
Curriculum coverage and mismatch are
\begin{equation}
\begin{aligned}
\mathrm{CCR}_k
&=\frac{|\mathcal{C}_k\cap\mathcal{E}_k|}
{|\mathcal{E}_k|},\\
\mathrm{CMR}_k
&=\frac{|\mathcal{C}_k\setminus\mathcal{E}_k|}
{|\mathcal{C}_k|}.
\end{aligned}
\end{equation}
These metrics measure alignment with the eligibility rule.

Table~\ref{tab:ablation} shows that SWE-Growth achieves
the highest mean performance on both benchmarks in
both rounds.
In Round~1, SFT only performs better than GRPO only
on SWE-bench Pro, but neither matches the full
method's cross-domain performance, supporting their
complementary roles.
By Round~2, GRPO only approaches the full method
while SFT only lags behind, suggesting a greater
marginal role for reinforcement learning after
the initial round of combined training. Task selection also matters.
Despite collecting fresh post-SFT rollouts,
No refresh retains stale task selections and
misses newly eligible learning opportunities.
Fixed-pair gap exhibits larger discrepancies in
both rounds, indicating that differences between
frozen models do not reliably track the current
Student's needs.
Their lower benchmark scores support refreshing
task selection against the evolving Student instead of merely collecting new rollouts.
\begin{figure}[t]
\centering
\includegraphics[width=\columnwidth]
{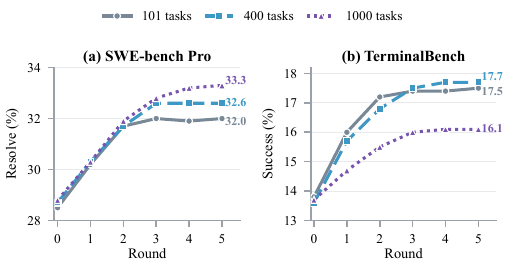}
\caption{
Performance over iterative SWE-Growth rounds at different
source-task scales.
}
\label{fig:scaling_rounds}
\end{figure}

\subsection{Scaling and saturation across self-improvement rounds.}
We study how SWE-Growth scales with the number of source tasks
and how many rounds of iterative training continue to provide
measurable gains. 
Starting from the same Student and Teacher, we run SWE-Growth
with $N=101$ source tasks and two enlarged sets of $N=400$ and $N=1000$ tasks.
For each setting, the Injector and Student continue from their
previous-round checkpoints, while the Teacher remains frozen.
We determine when to stop training using an independent
development set, separate from both evaluation benchmarks.
Training stops when development performance plateaus.

Figure~\ref{fig:scaling_rounds} shows that larger source-task pools support
higher observed in-domain scores over additional
training rounds: SWE-bench Pro reaches 32.0\%,
32.6\%, and 33.3\% with 101, 400, and 1000 tasks,
respectively.
Cross-domain gains are not monotonic:
TerminalBench reaches 17.5\%, 17.7\%, and 16.1\%.
Thus, scaling SWE-specific self-play improves
in-domain repair without ensuring further
cross-domain gains.
\section{Conclusion}

We introduced \textsc{SWE-Growth}, which uses task-level
Teacher--Student gaps to guide staged SFT--GRPO learning
and Injector updates.
Starting from 101 seed issues, it reaches 31.7\%
on SWE-bench Pro and 17.2\% on TerminalBench 2.1,
outperforming Teacher-augmented self-play.
Results with a weaker Teacher further highlight
the value of task-level complementarity.

\newpage
\section*{Limitations}
Our study covers a limited set of Teacher configurations.
The effects of Teacher capability and demonstration
format on Student improvement warrant further study.
We have not systematically compared the strengths
and limitations of task-level Teacher--Student gap
selection and token-level Teacher supervision.
These mechanisms may be complementary, and combining
them could yield further improvements.
Moreover, selecting tasks based on a frozen Teacher's
success may exclude tasks that the Student could learn
through other forms of supervision or exploration.

Our curriculum uses empirical Teacher--Student success
gaps estimated from a small number of rollouts
($G=4$ and $G=8$ per agent).
These estimates are sensitive to sampling variability:
a single rollout outcome can change a task's eligibility,
particularly near the selection threshold.
The gap therefore serves as a practical curriculum
selection signal rather than a precise measure of
relative task-solving ability.
Evaluating selection stability across independent
rollout samples and incorporating uncertainty into
task selection are directions for future work.

Our evaluation covers repository-level code repair
and terminal-based tasks.
Generalization to other domains remains to be established.

\section*{Ethical Statement}
SWE-Growth is intended to improve software repair,
but automated bug injection could be misused to
introduce defects, and agent-generated code may
cause unintended changes when executed.
Its use should therefore be restricted to authorized,
isolated environments with appropriate access controls.
Passing tests does not guarantee patch correctness
or security; generated repairs require human review
before deployment.
Repository data and repair trajectories may also
contain sensitive information or licensed code,
requiring careful handling, particularly when
using external Teacher services or releasing
training artifacts.
Finally, iterative generation and training incur
computational and environmental costs that should
be considered when scaling the framework.
\newpage
\bibliography{custom}

\appendix
\section{Additional Related Work}
\label{sec:add_related_work}

\paragraph{Adaptive curricula and self-play for language models.}
Prior work frames adaptive task generation as a curriculum-design
problem.
PAIRED generates tasks from the relative performance of two agents
\citep{NEURIPS2020_985e9a46}; Replay-Guided Adversarial Environment
Design selectively replays high-value tasks
\citep{NEURIPS2021_0e915db6}, while ACCEL modifies useful tasks to
track the learner's changing capability frontier
\citep{pmlr-v162-parker-holder22a}.
Related work shows that
irreducible high-regret tasks can cause learning to stagnate
\citep{NEURIPS2024_1d0ed12c,NEURIPS2025_45d58239,pmlr-v235-beukman24a}.
Language-model systems extend these ideas through role-based self-play:
Absolute Zero and R-Zero train task-proposal and solution roles
\citep{NEURIPS2025_9837dc00,huang2026rzeroselfevolvingreasoningllm};
Socratic-Zero co-evolves Teacher, Solver, and Generator agents
\citep{wang2025socraticzerobootstrappingreasoning}; SPIRAL, SPC, SPICE,
and J-Zero explore zero-sum interaction, adversarial criticism,
corpus-grounded task generation, and Challenger--Solver--Judge
co-evolution, respectively
\citep{liu2026spiralselfplayzerosumgames,NEURIPS2025_cb7baa00,liu2025spiceselfplaycorpusenvironments,chu2026jzerounifiedchallengersolverjudgecoevolution}.
Complementarily, DataEnvGym develops data-generation agents whose
teaching data adapts to feedback from the current Student
\citep{khan2025dataenvgym}, while SOAR rewards task generation through
measured Student learning progress
~\citep{sundaram2026teaching}.
However, adapting task difficulty or measuring
learning progress does not by itself determine
whether a generated repair task should supply
demonstrations or group-relative learning signals.
SWE-Growth makes these learning roles explicit
and uses their eligibility to reward the Injector.
Student updates therefore change both the use of
existing tasks and the feedback guiding subsequent
task generation.
\paragraph{Capability-calibrated task generation}
Capability differences provide a practical signal
for constructing challenging but solvable tasks.
AgentFrontier combines iterative question refinement
with calibration between a less knowledgeable model
and a more capable, tool-augmented agent, retaining
frontier questions that only the latter
solves~\citep{chen2025agentfrontier}.
CalibForge uses solver outcomes and trajectories
to revise executable terminal tasks until they
exhibit cross-solver disagreement or a designated
strong-pass/weak-fail relation~\citep{meng2026calibforge}.
These approaches show that task validity alone
does not establish training value and that solver
behavior can guide both filtering and task revision.
However, construction-time calibration does not
by itself specify how training eligibility should
change after the target Student is updated.
SWE-Growth evaluates task-level gaps against the
current Student, uses reliable Teacher successes
to select SFT demonstrations, and reassesses tasks
after SFT for mixed-outcome GRPO training.
The resulting eligibility also guides Injector
updates, connecting capability-based selection
with the Student's evolving learning process.
\paragraph{Teacher guidance and Student self-learning.}
Existing approaches obtain learning signals from
external Teachers or from the Student's own
attempts and feedback.
For external guidance, LUFFY combines off-policy
demonstrations with on-policy rollouts through
mixed-policy GRPO, while ReLIFT interleaves
reinforcement learning with fine-tuning on
solutions to challenging questions
~\citep{yan2026learning,ma2026learning}.
On-policy distillation supervises Student-generated
sequences using Teacher distributions, reducing
training--inference distribution mismatch
~\citep{agarwal2024policy}.
KDRL combines reverse-KL distillation with
reward-based optimization~\citep{xu2025kdrl},
while RSTG selectively distills Teacher guidance
into all-fail groups and supplements it with
successful Teacher trajectories~\citep{han2026distill}.
For Student self-learning, GRESO uses reward
history to selectively avoid uninformative
rollouts~\citep{zheng2026act}.
RL-ZVP shapes advantages using entropy, while
CPO uses disagreement between reference-guided
and vanilla policies~\citep{le2026no,xu2026beyond}.
SDPO constructs an evolving, feedback-conditioned
self-teacher~\citep{hubotter2026reinforcement},
and SRPO routes successful responses to GRPO
and failed responses with available teaching
signals to self-distillation~\citep{li2026unifying}.
However, the usefulness of guidance depends on
more than the Student's failure.
For external distillation, a Teacher's overall
strength does not guarantee reliable guidance
on every task.
For self-distillation, informative supervision
depends on the model's ability to use the
available feedback for self-correction.
These considerations motivate our use of
execution-verified, task-level Teacher--Student
gaps to select demonstrations, followed by
post-SFT reassessment for GRPO.
SWE-Growth uses execution-verified
Teacher--current Student gaps to select SFT
demonstrations, then reassesses Student outcomes
and routes mixed groups with a positive gap to GRPO.
This connects verified external guidance with
learning from the Student's own attempts, while
the resulting task eligibility also guides
subsequent Injector updates.

\section{Baselines, Evaluation And Training details}
\label{training_details}
\paragraph{Baselines}
Table~\ref{main_table} separates controlled training
comparisons from external model references.
Our implementations of SWE-MeM SFT + RL, Self-Play SWE-RL and Teacher
Distillation use Qwen3-Coder-30B-A3B-Instruct as the
Student backbone and Mini-SWE-Agent as the default scaffold.
Self-Play SWE-RL provides the Student-only iterative
self-play comparison. Teacher Distillation trains the
Student on Teacher demonstrations from 6805 initial tasks collected from SWE-smith and SWE-mirror initial issues. SWE-MeM constructs its SFT and RL data from SWE-ReBench
and SWE-Gym. For SFT, it generates three rollouts per
problem using Qwen3-Coder-30B-A3B-Instruct for
task-solving actions; GPT-5.1 is used to synthesize
memory-management supervision. After filtering, the
SFT set contains 8,728 trajectory samples from 5,867
unique problems. For RL, SWE-MeM removes instances
whose reference patches fail the associated tests and
selects problems with intermediate success rates under
best-of-$n$ verification, yielding 2,194 training
instances.
\subsection{Training details.}
\label{Training details}
\paragraph{Round-0 Source tasks and Initialization}
We initialize the Student Fixer and Injector from
Qwen3-Coder-30B-A3B-Instruct, while keeping the
Teacher frozen. The initial self-play pool contains
101 distinct resolved issues from 92 repositories.
It covers Python (31), JavaScript (29), TypeScript
(27), Go (12), C (1), and Java (1). For every source
issue, the original tests fail, the developer patch
applies, and the tests pass after the patch. We use
the developer patch to restore a clean, repaired
repository, then let the Injector introduce a new
bug. The Student receives the resulting repair task
without access to the developer patch or injection
patch. Execution-validated synthetic tasks provide
the experience used for subsequent Fixer updates.

Both trainable policies use Qwen3-Coder-30B-A3B-Instruct
as their backbone. To initialize the Student Fixer,
we supervise it on successful Kimi-K3 execution
trajectories for 12 SFT steps. We then identify bugs
that the initialized Student cannot solve. For these
bugs, we reverse successful repair paths to construct
bug-injection examples and train the initial Injector
on them for another 12 SFT steps. The resulting Student
and Injector checkpoints start the iterative self-play
rounds.
\paragraph{Round-1 Fixer training details.}
The Fixer uses Qwen3-Coder-30B-A3B-Instruct as its
backbone. In this round, SFT starts from the
preceding Student adapter and trains on 36 records:
20 verified Teacher demonstrations and 16 verified
replay examples. The records represent 36 synthetic
instances from 20 distinct source-instance IDs.
Only the target assistant decision contributes to the
cross-entropy loss; preceding messages and tool outputs
provide context. SFT runs for one epoch on four GPUs
with distributed data parallelism, a per-GPU batch size
of one, gradient accumulation of two, and a partial
final batch. It uses a maximum training sequence length
of 65{,}536 tokens, a learning rate of $5\times10^{-8}$,
one warmup step, and a rank-8 LoRA adapter targeting
the query, key, value, and output projections. All
36 records are retained, yielding five optimizer updates.

Offline GRPO starts from this SFT adapter.
Its training data contain nine mixed-outcome frontier
groups and six retention-anchor groups.

\paragraph{Decision records and loss masking.}
A decision record consists of a retained agent action
and its preceding interaction history.
For each interaction turn, we retain the final agent
action as the prediction target. This action may involve
file inspection, editing, testing, final verification,
or submission. A single rollout trajectory can therefore
contribute multiple decision records across its turns.
Each trajectory contributes decision records containing
retained actions and their preceding histories.
The nine frontier groups and six anchor groups yield
911 such records.

Within each decision record, only tokens belonging to the
retained action contribute to the optimization objective;
preceding messages and tool outputs serve as conditioning
context and are excluded from the loss.
Frontier records contribute to the GRPO objective, whereas
retention-anchor records contribute only to the anchor
KL regularization described in Appendix~B.2. Each optimizer update uses
one intact frontier group and one anchor group; one
epoch therefore yields nine updates. Anchor groups
contribute only a KL penalty against the frozen SFT
adapter and are cycled when needed. GRPO uses a
learning rate of $5\times10^{-8}$, a maximum training
sequence length of 65{,}536 tokens, clipping parameter
$\epsilon=0.2$, a reference-KL coefficient of $0.02$,
an anchor-KL coefficient of $0.2$, and gradient-norm
clipping at $0.5$.

\paragraph{Training--evaluation separation.}
We construct the source-task pool independently of
SWE-bench Pro and Terminal-Bench.
Training and evaluation repositories are disjoint:
none of the source tasks originate from repositories
included in either benchmark.
All synthetic repair tasks are generated exclusively
from these training repositories.
Benchmark issues, patches, and tests are not used for
bug generation or training.

\paragraph{GRPO Objective Details}
\label{app:grpo_details}

For each task, let $r_g=r(t_b,p_g)$ denote the binary
repair reward of rollout $g$.
The group-relative advantage is
\begin{equation}
\begin{aligned}
\bar r &= \frac{1}{G}\sum_{g=1}^{G}r_g,\\
\sigma_r &=
\sqrt{\frac{1}{G}\sum_{g=1}^{G}(r_g-\bar r)^2},\\
\widehat A_g &=
\frac{r_g-\bar r}{\sigma_r+\epsilon_{\mathrm{adv}}},
\end{aligned}
\end{equation}
where $\epsilon_{\mathrm{adv}}>0$ ensures numerical
stability.
The group mean provides the baseline; no learned
value network is used.

For rollout $g$, let
$\{(h_{g,d},a_{g,d})\}_{d=1}^{D_g}$
denote its retained decision records, constructed as
described in Appendix~B.1.
Here, $h_{g,d}$ is the preceding interaction history
and $a_{g,d}$ is the final agent action retained
from the corresponding interaction turn.
A rollout may contribute multiple decision records.
All records from rollout $g$ share its group-relative
advantage $\widehat{A}_g$.

Let $c_{g,d,j}=(h_{g,d},a_{g,d,<j})$ denote the
conditioning context for action token $j$.
The importance ratio is
\begin{equation}
\rho_{g,d,j}(\theta)
=
\frac{\pi_\theta(a_{g,d,j}\mid c_{g,d,j})}
     {\pi_{\mathrm{old}}(a_{g,d,j}\mid c_{g,d,j})},
\end{equation}
where $\pi_{\mathrm{old}}$ is the frozen
rollout-generating Student that initializes the
corresponding GRPO training stage.

We use the frozen post-SFT Student $\pi_{\mathrm{KL}}$
as the KL reference, distinct from the Teacher used
for curriculum selection.
For $c_{g,d,j}=(h_{g,d},a_{g,d,<j})$, define
\begin{equation}
\begin{aligned}
u_{g,d,j}(\theta)
&=
\frac{\pi_{\mathrm{KL}}(a_{g,d,j}\mid c_{g,d,j})}
{\pi_\theta(a_{g,d,j}\mid c_{g,d,j})},\\
K_{g,d,j}(\theta)
&=u_{g,d,j}(\theta)
-\log u_{g,d,j}(\theta)-1.
\end{aligned}
\end{equation}
This sampled penalty targets
$D_{\mathrm{KL}}(\pi_\theta\Vert\pi_{\mathrm{KL}})$.
Its expectation equals this forward KL when tokens
are sampled from $\pi_\theta$; during optimization,
we evaluate it on the fixed rollout tokens generated
by $\pi_{\mathrm{old}}$.

Let $D_g$ be the number of retained assistant
decisions in rollout $g$.
The mask $m_{g,d,j}\in\{0,1\}$ equals one only
for valid action tokens selected for optimization.
Conditioning-only tokens, padding, and excluded
tokens receive zero weight.
The number of optimized action tokens in rollout
$g$ is
\begin{equation}
N_g =
\sum_{d=1}^{D_g}
\sum_{j=1}^{|a_{g,d}|}m_{g,d,j}.
\end{equation}
The rollout loss is normalized by $N_g$.
The frontier-group loss averages these
token-normalized losses equally across its $G$
rollouts.
The frontier-group loss is defined as follows.
\par
\begingroup
\small
\begin{equation}
\begin{alignedat}{1}
&\mathcal{L}_g(\theta)
= -\frac{1}{N_g}
\sum_{d=1}^{D_g}
\sum_{j=1}^{|a_{g,d}|}
m_{g,d,j}
\Bigl[
\phi(\rho_{g,d,j}(\theta),\widehat{A}_g)
\\
&\qquad -\beta K_{g,d,j}(\theta)\Bigr],
\\
&\mathcal{L}_{\mathrm{GRPO}}(\theta)
= \mathbb{E}_{\mathcal{G}}\!\left[
\frac{1}{G}\sum_{g=1}^{G}\mathcal{L}_g(\theta)
\right].
\end{alignedat}
\end{equation}
\par
\endgroup

Here, $\phi$ is the clipped surrogate defined in
Equation~(10), and the expectation is over eligible
post-SFT rollout groups.
Only retained action tokens contribute to the loss;
preceding messages and tool outputs provide context.
Each trajectory receives equal weight in the
group-level loss, regardless of its number of
decision records or valid action tokens.
We use $\epsilon=0.2$ and $\beta=0.02$.

Each update also includes a retention-anchor group,
which receives only KL regularization against the
frozen SFT adapter, without a reward-advantage term.
Writing its aggregated KL penalty as
$\mathcal L_{\mathrm{anchor}}$, the total loss is
\begin{equation}
\mathcal L_{\mathrm{total}}
=
\mathcal L_{\mathrm{GRPO}}
+
\beta_{\mathrm{anchor}}\mathcal L_{\mathrm{anchor}},
\end{equation}
where $\beta_{\mathrm{anchor}}=0.2$.

\begin{table*}[t]
\centering
\small
\caption{Arena evaluation configuration. The values describe the
registered harnesses used in the reported SWE-Growth runs; task-specific
verifier commands may impose additional limits.}
\label{tab:arena-eval-config}
\begin{tabular}{p{0.19\textwidth}p{0.36\textwidth}p{0.36\textwidth}}
\toprule
Item & SWE-bench Pro & Terminal-Bench 2 \\
\midrule
Benchmark and harness
& \texttt{swebench-pro@1.2.2};

& \texttt{tb21-terminus2-full@2.1.1};
  \texttt{terminus-2-tb21@1.0.10} \\

Implementation
& OpenHands \texttt{1.14.0-internal.1};

& Terminus \texttt{2.0.0}; \\

Agent interaction
& Native tool calls, including \texttt{ThinkTool} and
  \texttt{FinishTool}; concise offline task prompt; no conversation
  condenser; one tool call at a time
& Terminus 2 task prompt and JSON-formatted terminal actions executed
  through a persistent \texttt{tmux} session; no artificial turn cap \\

Context and output
& Input context limit 262{,}144 tokens; configured maximum output
  128{,}000 tokens
& Context limit 262{,}144 tokens; no separate output-token cap
  specified by the registered harness \\

Sampling
& Temperature $1.0$, top-$p$ $0.95$, top-$k$ $50$;
  one sample per task
& Temperature $1.0$, top-$p$ $0.95$, top-$k$ $50$;
  one sample per task \\

Time limits
& Agent harness TTL 7{,}200\,s; environment TTL 14{,}400\,s;
  reward TTL 7{,}200\,s
& Agent budget 14{,}400\,s; environment TTL 29{,}400\,s;
  reward TTL 7{,}200\,s \\

Tool and test limits
& No single uniform tool-call or test-command timeout is specified in
  the registered harness; the grader runs each task's prescribed test
  script within its reward budget
& Terminal commands run through \texttt{tmux}; the prompt recommends
  waits of at most 60\,s and polling for long commands, not a 60\,s
  command deadline; the task verifier runs within its reward budget \\

Submission and grading
& The harness extracts a source patch from the final repository-tree
  difference relative to the task baseline. The grader applies it to
  the task repository and runs the prescribed tests
& No patch submission contract; the task-specific verifier grades the
  final environment state \\

Retries and recovery
& Up to two fresh completion-correction conversations on the same
  workspace when a valid native \texttt{FinishAction} is missing;
  failed patches are not salvaged. Arena infrastructure retries are
  recorded separately
& Up to six transport retries per model request; partial trajectories
  are retained on agent timeout. Arena infrastructure retries are
  recorded separately \\

Score and exclusions
& Binary task reward, averaged across tasks; Arena's aggregate treats
  failed tasks as zero. Paired analyses exclude tasks without valid
  completed results from either system
& Binary task reward, averaged across tasks; Arena's aggregate treats
  failed tasks as zero. Paired analyses use the common completed subset \\
\bottomrule
\end{tabular}
\end{table*}
\paragraph{Evaluation protocol.}
\label{evaluation_details}
We evaluate each checkpoint under three evaluation
seeds, using one model sample per task per seed.
Table~2 reports the mean and standard deviation
of the three benchmark-level scores, without
best-of-seed selection. We assess significance
using a paired t-test on seed-matched benchmark
scores, comparing Round-2 SWE-Growth with Round-2
Self-Play SWE-RL + Teacher on each benchmark.
Infrastructure retries and harness-level recovery are tracked separately
from model sampling. For paired comparisons, we include only tasks
completed by both systems and report excluded harness failures and
missing results separately. Table~\ref{tab:arena-eval-config} gives
the exact harness versions and execution limits.


\begin{table*}
\centering
\small
\setlength{\tabcolsep}{4pt}
\resizebox{\textwidth}{!}{%
\begin{tabular}{llcccccc}
\toprule
\multicolumn{2}{c}{Teacher configuration}
& \multicolumn{2}{c}{Teacher SWE-bench Pro (\%)}
& \multicolumn{2}{c}{Student SWE-bench Pro (\%)}
& \multicolumn{2}{c}{Cost (USD / 1M tokens)} \\
\cmidrule(lr){1-2}
\cmidrule(lr){3-4}
\cmidrule(lr){5-6}
\cmidrule(lr){7-8}
Teacher A & Teacher B
& A & B
& Round 1 & Round 2
& A (input/output) & B (input/output) \\
\midrule
Qwen3.6-35B-A3B & ---
& 38.5 & --- & 30.2 & 31.7 & 0.25/1.50 & --- \\

Qwen3.8-35B-A3B & ---
& 44.6 & --- & 30.4 & 31.8 & 0.33/1.85 & --- \\

DeepSeek-V4-Pro & ---
& 51.2 & --- & 29.9 & 31.1 & 0.66/1.98 & --- \\
\midrule
Qwen3.6-35B-A3B & Qwen3.8-35B-A3B
& 38.5 & 44.6 & 30.5 & 32.0 & 0.25/1.50 & 0.33/1.85 \\

Qwen3.6-35B-A3B & DeepSeek-V4-Pro
& 38.5 & 51.2 & 30.2 & 31.4 & 0.25/1.50 & 0.66/1.98 \\
\bottomrule
\end{tabular}%
}
\caption{
Teacher selection and Student performance.
Teacher scores are measured under the same evaluation protocol.
Student scores are reported after training rounds 1 and 2.
Input and output token prices are listed separately for each
Teacher; unavailable entries are marked with dashes.
}
\label{tab:teacher_selection}
\end{table*}
\section{More experiments}
\subsection{More Experiments on Teacher Selection, Cost, and Student Performance}
\label{more_teachers}
Table~\ref{tab:teacher_selection} compares three single-Teacher
and two multi-Teacher configurations.
Using Qwen3.8 instead of Qwen3.6 slightly improves Student
performance, while combining the two achieves the highest
scores of 30.5\% and 32.0\% after rounds 1 and 2.
We use Qwen3.6 alone by default because it provides
competitive Student performance at the lowest listed
input and output token prices, without requiring a second
Teacher.

Notably, stronger Teacher performance does not necessarily
translate into greater Student improvement.
Although DeepSeek-V4-Pro achieves the highest Teacher
score (51.2\%), the Student trained with it reaches only
31.1\% in round 2, compared with 31.7\% using Qwen3.6.
Combining DeepSeek with Qwen3.6 also falls below Qwen3.6
alone (31.4\% versus 31.7\%).
These results suggest that standalone repair accuracy
is insufficient for selecting an effective Teacher.
Differences in demonstration content, output format,
or compatibility with the Student may contribute,
but identifying the underlying causes requires further
investigation.
\begin{table}[t]
\centering
\small
\setlength{\tabcolsep}{3pt}
\begin{tabular}{lcc}
\toprule
Metric & Async. & Sync. \\
\midrule
End-to-end time (h) $\downarrow$ & 2.63 & 5.59 \\
Time to first SFT (h) $\downarrow$ & 1.03 & 1.74 \\
Evaluated tasks/h $\uparrow$ & 24.86 & 15.15 \\
\midrule
Evaluated tasks & 52 & 52 \\
SFT / GRPO tasks & 11/8 & 11/8 \\
Evaluated rollouts (SFT/GRPO stages) & 132/64 & 132/64 \\
Rollout tokens (M) & 98.369 & 96.172 \\
\bottomrule
\end{tabular}
\caption{
Synchronous and asynchronous execution with the same
worker allocation and task budget.
Evaluated tasks count candidates whose repair evaluation
and routing are complete, including those not selected
for training.
Throughput is measured over the generation and evaluation
phase.
Rollout tokens include Injector, Student, and Teacher
generation.
}
\label{tab:async_efficiency}
\end{table}
\subsection{Asynchronous execution efficiency.}
\label{async}
Table~\ref{tab:async_efficiency} shows that asynchronous
execution reduces end-to-end time from 5.59 to 2.63 hours,
achieving a $2.13\times$ speedup under the same resource
allocation.
SFT begins 40.8\% earlier, while evaluation throughput
increases from 15.15 to 24.86 tasks per hour (64.1\%).
Both settings evaluate 52 tasks and use the same numbers
of selected training tasks and optimization updates.
Rollout-token usage increases by only 2.3\%.
These results show that asynchronous execution reduces
elapsed training time without reducing the amount of
training performed or substantially increasing
rollout-token usage.

\begin{figure}[t]
    \centering
    \includegraphics[width=\columnwidth]
    {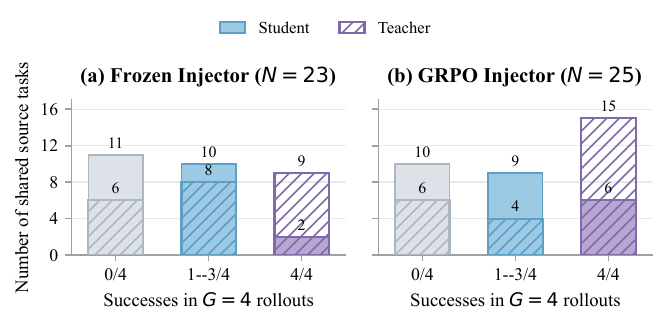}
    \caption{
    Downstream repair outcomes for tasks generated by the frozen
    and GRPO-trained Injectors.
    Solid bars show Student outcomes, while hatched bars show
    Teacher outcomes.
    }
    \label{fig:injector_outcomes}
\end{figure}
\subsection{Ablation study of Injectors and bug difficulty analysis}
We compare a frozen Injector with the GRPO-trained Injector by
evaluating their generated bugs using the same Student and frozen
Teacher, with $G=4$ repair attempts per task.
As shown in Figure~\ref{fig:injector_outcomes}, Injector training
changes the downstream difficulty distribution.
The proportion of tasks solved in all four attempts increases from
8.7\% to 24.0\% for the Student and from 39.1\% to 60.0\% for the
Teacher.
The clearest change appears among tasks with mixed Student outcomes served as GRPO candidates:
the Teacher achieves a higher success count on all 9 such tasks
generated by the trained Injector, compared with 6 of 10 under the
frozen Injector.
These results suggest that the gain from Injector training comes from
reshaping the generated-task distribution toward bugs that provide
Student rollout variation together with a reliable Teacher reference,
making them better suited to the Teacher-guided GRPO curriculum.

\subsection{Case analysis of injected bugs by frozen and GRPO-trained injector}
Figure~\ref{fig:injector-case-study} shows how training the injector can
improve the usefulness of generated bugs for downstream Fixer learning.
In React Redux, the GRPO-trained injector produced a validated bug that
remained challenging for the student (2/4 solved), while teacher success
increased from 1/4 to 4/4. The teacher--student gap consequently increased
from $-1/4$ to $+2/4$, creating a clearer opportunity for learning from
teacher behavior. Auth0 illustrates a complementary benefit: student
success increased from 0/4 to 2/4 while teacher success remained at 3/4.
The trained-injector bug therefore yielded a mixed student group, providing
both successful and unsuccessful trajectories for group-relative learning.
Together, these cases illustrate that injector training can improve
downstream data utility by increasing teacher advantage or by making
student outcomes more informative. 
\begin{figure*}[t]
\centering
\begin{minipage}[t]{0.56\textwidth}
\small
\textbf{(a) React Redux: representative injected changes}\\[3pt]
\fbox{\begin{minipage}{0.95\linewidth}
\textbf{Frozen injector}\\[-2pt]
{\ttfamily\scriptsize
\textcolor{red}{- const previousState = useMemo(...)}\\
\textcolor{blue}{+ const previousStore = usePrevious(store)}\\
\textcolor{blue}{+ const previousState = useMemo(...)}}
\end{minipage}}\\[6pt]
\fbox{\begin{minipage}{0.95\linewidth}
\textbf{GRPO-trained injector}\\[-2pt]
{\ttfamily\scriptsize
\textcolor{red}{- if (previousState !== store.getState()) \{}\\
\textcolor{red}{- \ \ subscription.notifyNestedSubs()}\\
\textcolor{red}{- \}}}
\end{minipage}}
\end{minipage}
\hfill
\begin{minipage}[t]{0.40\textwidth}
\small
\textbf{(b) Fixer success, out of four rollouts}\\[3pt]
\begin{tabular}{llcc}
\hline
Source & Injector & Student & Teacher\\
\hline
React Redux & Frozen & 2/4 & 1/4\\
            & GRPO   & 2/4 & 4/4\\
\hline
Auth0       & Frozen & 0/4 & 3/4\\
            & GRPO   & 2/4 & 3/4\\
\hline
\end{tabular}
\end{minipage}
\caption{\textbf{Source-matched injector case studies.}
For React Redux, both injected bugs independently passed the validity check
(injected baseline reward 0; reference repair reward 1).
With matched Fixers and evaluation settings, the GRPO-injector bug preserves
student difficulty (2/4 solved) while teacher success increases from 1/4 to 4/4.
In the earlier Auth0 evaluation, the GRPO-injector bug yields a mixed student
group (2/4 versus 0/4), while teacher success remains 3/4.}
\label{fig:injector-case-study}
\end{figure*}

\begin{table}[t]
\centering
\small
\setlength{\tabcolsep}{3pt}
\begin{tabular}{lcc}
\toprule
Resource & SP + T & SWE-Growth \\
\midrule
Injection attempts       & 356 & 364 \\
Teacher repair attempts  & 104 & 110 \\
Teacher input tokens (M) & 55.438 & 56.242 \\
Teacher output tokens (M)& 0.778 & 0.756 \\
Total rollout tokens (M) & 100.539 & 98.369 \\
Training Time(min)       & 41.4 & 38.6 \\
\bottomrule
\end{tabular}
\caption{
Cumulative resource usage across Rounds 0--2.
SP + T denotes Self-Play SWE-RL + Teacher.
Total rollout tokens include Injector, Student,
and Teacher generation; loss tokens count only
tokens contributing to optimization.
}
\label{tab:teacher_budget_usage}
\end{table}

\begin{table}[t]
\centering
\small
\setlength{\tabcolsep}{6pt}
\begin{tabular}{lcc}
\toprule
Round & Pro (\%) & TB (\%) \\
\midrule
\multicolumn{3}{l}{\textit{Self-Play SWE-RL}} \\
0 & $28.3 \pm 1.3$ & $11.8 \pm 1.7$ \\
1 & $28.8 \pm 1.7$ & $12.7 \pm 1.9$ \\
2 & $29.2 \pm 1.7$ & $13.3 \pm 2.2$ \\
\midrule
\multicolumn{3}{l}{\textit{Self-Play SWE-RL + Teacher}} \\
0 & $28.6 \pm 1.6$ & $12.9 \pm 1.8$ \\
1 & $29.3 \pm 1.5$ & $13.9 \pm 1.7$ \\
2 & $30.1 \pm 1.9$ & $14.8 \pm 2.1$ \\
\midrule
\multicolumn{3}{l}{\textsc{SWE-Growth}} \\
0 & $28.6 \pm 1.2$ & $14.0 \pm 1.9$ \\
1 & $30.3 \pm 1.6$ & $16.0 \pm 1.8$ \\
2 & $\mathbf{32.0 \pm 1.8}$ & $\mathbf{17.4 \pm 1.9}$ \\
\bottomrule
\end{tabular}
\caption{
Performance with $G=8$ on SWE-bench Pro (Pro)
and TerminalBench 2.1 (TB).
Each method follows its own training trajectory.
Scores are mean $\pm$ standard deviation
across three evaluation seeds.
}
\label{tab:g8_results}
\end{table}
Table~\ref{tab:g8_results} shows that SWE-Growth
maintains its advantage with $G=8$, reaching
32.0\% on SWE-bench Pro and 17.4\% on TerminalBench
in Round 2.
It exceeds Student-only self-play by 2.8 and
4.1 percentage points, and Teacher-augmented
self-play by 1.9 and 2.6 points, respectively.
Compared with its $G=4$ results, SWE-Growth
achieves slightly higher mean scores
(+0.3 and +0.2 points).
These results indicate that its advantage persists
under a larger rollout budget, while doubling
the group size yields only modest additional
gains in this setting.
\subsection{Effect of Rollout Group Size}
\label{sec:rollout_group_size}

We examine whether increasing the rollout group
size changes training performance and the availability
of useful learning signals.

\paragraph{Training performance with $G=8$.}

We increase the number of repair attempts per task
from four to eight for both the Student and Teacher.
The SFT success-rate threshold remains unchanged:
the Student must fail all eight attempts, while
the Teacher must succeed in at least four.
We compare SWE-Growth with Self-Play SWE-RL and
Self-Play SWE-RL + Teacher over Rounds 0--2.
Each method continues from its own preceding-round
checkpoint under the same configured per-round
resource limits.
Table~\ref{tab:g8_results} reports the results.

\begin{table}[t]
\centering
\small
\setlength{\tabcolsep}{5pt}
\begin{tabular}{lcc}
\toprule
Task category & $G=4$ & $G=8$ \\
\midrule
Student all-fail ($s_S=0$)
& 34.9\% & 32.1\% \\
Student mixed ($0<s_S<1$)
& 49.4\% & 55.3\% \\
Student all-pass ($s_S=1$)
& 15.7\% & 12.6\% \\
\midrule
$s_S=0,\ s_T\geq 1/2$
& 23.3\% & 22.8\% \\
$0<s_S<s_T$
& 36.7\% & 35.5\% \\
\bottomrule
\end{tabular}
\caption{
Group composition with fixed checkpoints initialzed from the same 101 seed issues.
The first three rows partition the task pool;
the final two report Teacher-gap subsets.
The Teacher threshold $s_T\geq 1/2$ requires
at least two successes for $G=4$ and four for $G=8$.
}
\label{tab:group_size_composition}
\end{table}

Table~\ref{tab:g8_results} shows that SWE-Growth
maintains its advantage with $G=8$, reaching
32.0\% on SWE-bench Pro and 17.4\% on TerminalBench
in Round 2.
It exceeds Student-only self-play by 2.8 and
4.1 percentage points, and Teacher-augmented
self-play by 1.9 and 2.6 points, respectively.
Compared with its $G=4$ results, SWE-Growth
achieves slightly higher mean scores
(+0.3 and +0.2 points).
These results indicate that its advantage persists
under a larger rollout budget, while doubling
the group size yields only modest additional
gains in this setting.

\paragraph{Group composition with $G=4$ and $G=8$.}
To isolate the effect of additional sampling,
we hold the task pool and Student and Teacher
checkpoints fixed.
Each model makes eight attempts per task;
the first four form the corresponding $G=4$ group.
Table~\ref{tab:group_size_composition} reports
Student all-fail, mixed, and all-pass proportions,
alongside two Teacher-gap subsets:
Student all-fail tasks with Teacher success rate
at least $1/2$, and mixed Student tasks with
a higher Teacher success rate.
All proportions use the same denominator of
tasks with complete evaluations for both models.

Increasing $G$ from 4 to 8 reduces the proportions
of all-fail and all-pass Student groups, while
increasing mixed groups from 49.4\% to 55.3\%.
However, the two Teacher-gap subsets remain
similar in proportion: Student all-fail tasks
with reliable Teacher successes change from
23.3\% to 22.8\%, and mixed tasks with a positive
Teacher gap from 36.7\% to 35.5\%.
Thus, although additional sampling reveals more
mixed outcomes, the aggregate proportions of
tasks meeting our curriculum criteria remain
largely stable across the two group sizes.

\begin{table}[t]
\centering
\small
\setlength{\tabcolsep}{4pt}
\begin{tabular}{lccc}
\toprule
GRPO selection & Round & Pro (\%) & TB (\%) \\
\midrule
\textsc{SWE-Growth}
& 1 & 30.2 & 16.0 \\
Negative gap
& 1 & 28.1 & 10.5 \\
All mixed
& 1 & 29.9 & 15.2 \\
Random mixed (matched size)
& 1 & 29.1 & 14.5 \\
\midrule
\textsc{SWE-Growth}
& 2 & 31.7 & 17.2 \\
Negative gap
& 2 & 29.8 & 15.5 \\
All mixed
& 2 & 31.0 & 16.3 \\
Random mixed (matched size)
& 2 & 30.5 & 16.0\\
\bottomrule
\end{tabular}
\caption{
Single-round ablations of post-SFT GRPO selection.
Variants share the same within-round starting
checkpoint and SFT stage.
Round-2 variants inherit standard SWE-Growth
Round-1 training, not their own ablation histories.
Random mixed matches SWE-Growth's selected
group count within each round.
}
\label{tab:post_sft_gap_ablation}
\end{table}

\subsection{Additional Ablation Study}
\label{sec:additional_ablation}
\paragraph{Effect of selecting positive Teacher-Student Gap in GRPO}
We examine whether a positive Teacher--Student gap
helps select useful GRPO tasks beyond requiring
mixed Student outcomes.
All variants retain the same SFT stage and use
the same post-SFT candidate rollouts, differing
only in GRPO task selection.

Let $s_S$ and $s_T$ denote the post-SFT Student
and frozen Teacher success rates.
SWE-Growth selects tasks with $0<s_S<s_T$.
We compare three alternatives:
\textbf{Negative gap} selects mixed Student groups
with $s_T<s_S$;
\textbf{All mixed} selects all groups with
$0<s_S<1$, regardless of Teacher performance;
and \textbf{Random mixed (matched size)} uniformly
samples from these mixed groups to match
SWE-Growth's selected group count in each round.
The size-matched comparison separates the effect
of gap-based selection from that of training-pool size.

Table~\ref{tab:post_sft_gap_ablation} shows that
positive-gap selection achieves the highest scores
on both benchmarks in both rounds.
Compared with Random mixed under matched group
counts, SWE-Growth improves SWE-bench Pro by
1.1--1.2 percentage points and TerminalBench by
1.2--1.5 points, supporting the value of gap-based
selection beyond training-pool size.
Selecting all mixed groups also underperforms
SWE-Growth, while Negative gap yields the lowest
scores throughout.
Together, these results support using the
Teacher's task-level advantage to identify
productive GRPO tasks among those already
yielding mixed Student outcomes.

\begin{table}[t]
\centering
\small
\setlength{\tabcolsep}{4pt}
\resizebox{\columnwidth}{!}{%
\begin{tabular}{lccc}
\toprule
Method & Round & Pro (\%) & TB (\%) \\
\midrule
\textsc{SWE-Growth}
& 1 & \textbf{30.2} & \textbf{16.0} \\
SWE-Growth w/o warmup
& 1 & 29.4 & 13.9 \\
Self-Play SWE-RL + T w/o warmup
& 1 & 28.7 & 13.3 \\
\midrule
\textsc{SWE-Growth}
& 2 & \textbf{31.7} & \textbf{17.2} \\
SWE-Growth w/o warmup
& 2 & 30.5 & 15.8 \\
Self-Play SWE-RL + T w/o warmup
& 2 & 29.4 & 14.6 \\
\bottomrule
\end{tabular}%
}
\caption{
Effect of trajectory warmup on SWE-bench Pro
(Pro) and TerminalBench 2.1 (TB). T is short for Teacher.
}
\label{tab:warm_up}
\end{table}

\paragraph{Effect of Kimi-K3's warmup}
Table~\ref{tab:warm_up} shows that warmup improves
SWE-Growth's performance on both benchmarks,
but does not fully account for its advantage
over Teacher-augmented self-play.
Without warmup, SWE-Growth still outperforms
the corresponding baseline in both rounds,
reaching 30.5\% on SWE-bench Pro and 15.8\%
on TerminalBench in Round 2, compared with
29.4\% and 14.6\%, respectively.
Restoring warmup further improves these scores
by 1.2 and 1.4 percentage points.
These results support the benefit of SWE-Growth
beyond its trajectory initialization.
\subsection{Case Study}
\paragraph{Success compared with Frozen Teacher and Initial Student Backbone}
\begin{figure*}[t]
\centering

{\small
\textbf{Repair target:}
Convert overflow during \texttt{QPoint} construction
into \texttt{ValueError}.
\par}
\vspace{0.5em}

\begin{minipage}[t]{0.325\textwidth}
\vspace{0pt}
\begin{tcolorbox}[
  repaircard,
  title={Base Student},
  colframe=caseGray!65,
  colbacktitle=caseGray!8
]
{\scriptsize Qwen3-Coder-30B-A3B\par}
\smallskip
{\footnotesize
\textbf{Success: 0/3}
\hfill Sample 0: Fail\par}

\begin{lstlisting}[
  style=repaircase,
  escapeinside={(*@}{@*)}
]
try:
    x = int(x_str)
    y = int(y_str)
except OverflowError:
    raise ValueError(...)

(*@\textcolor{resultFail}{\bfseries return QPoint(x, y)}@*)
\end{lstlisting}

\tcblower
{\footnotesize
Guards integer parsing, but leaves
\texttt{QPoint} construction unprotected.
\par}
\end{tcolorbox}
\end{minipage}
\hfill
\begin{minipage}[t]{0.325\textwidth}
\vspace{0pt}
\begin{tcolorbox}[
  repaircard,
  title={Frozen Teacher},
  colframe=caseBlue!65,
  colbacktitle=caseBlue!8
]
{\scriptsize Qwen3.6-35B\par}
\smallskip
{\footnotesize
\textbf{Success: 2/3}
\hfill Sample 0: Pass\par}

\begin{lstlisting}[
  style=repaircase,
  escapeinside={(*@}{@*)}
]
(*@\textcolor{resultPass}{\bfseries try:}@*)
    return QPoint(x, y)
(*@\textcolor{resultPass}{\bfseries except OverflowError:}@*)
    (*@\textcolor{resultPass}{\bfseries raise ValueError(...)}@*)
\end{lstlisting}

\tcblower
{\footnotesize
Guards \texttt{QPoint} construction
and converts the overflow exception.
\par}
\end{tcolorbox}
\end{minipage}
\hfill
\begin{minipage}[t]{0.325\textwidth}
\vspace{0pt}
\begin{tcolorbox}[
  repaircard,
  title={SWE-Growth},
  colframe=caseGreen!75,
  colbacktitle=caseGreen!9
]
{\scriptsize Trained Qwen3-Coder-30B-A3B\par}
\smallskip
{\footnotesize
\textbf{Success: 3/3}
\hfill Sample 0: Pass\par}

\begin{lstlisting}[
  style=repaircase,
  escapeinside={(*@}{@*)}
]
(*@\textcolor{resultPass}{\bfseries try:}@*)
    return QPoint(x, y)
(*@\textcolor{resultPass}{\bfseries except Exception as e:}@*)
    (*@\textcolor{resultPass}{\bfseries raise ValueError(...)}@*) \
        (*@\textcolor{resultPass}{\bfseries from e}@*)
\end{lstlisting}

\tcblower
{\footnotesize
Guards the same operation with a broader
handler and preserves the exception cause.
\par}
\end{tcolorbox}
\end{minipage}

\caption{
A teacher-like repair pattern on Qutebrowser.
Success counts summarize three attempts per model;
code excerpts show abbreviated sample-0 patches.
Red marks the unguarded operation in the base
Student's patch; green marks the exception-handling
logic in the passing Teacher and SWE-Growth patches.
}
\label{fig:teacher_student_repair}
\end{figure*}

Figure~\ref{fig:teacher_student_repair}
illustrates how SWE-Growth can apply a repair
strategy also exhibited by the Teacher.
The base Student guards integer parsing but leaves
the subsequent \texttt{QPoint} construction unprotected,
missing the operation that can raise the overflow error.
Both the Teacher and SWE-Growth instead guard
this construction and convert the resulting exception
into the required \texttt{ValueError}.
SWE-Growth uses a broader exception handler
and preserves the original exception as its cause,
rather than reproducing the Teacher's patch verbatim.
Across three attempts, it succeeds three times,
compared with two Teacher successes and none
for the base Student.
This case is consistent with learning a Teacher-like
error-handling strategy, although patch similarity
alone does not establish direct transfer.

\paragraph{Failure compared with Frozen Teacher and Initial Student Backbone}
\begin{figure*}[t]
\centering

{\small
\textbf{Repair target:}
Call \texttt{add\_db\_name} during record expansion.
\par}
\vspace{0.5em}

\begin{minipage}[t]{0.49\textwidth}
\vspace{0pt}
\begin{tcolorbox}[
  repaircard,
  equal height group=openlibrarycomparison,
  title={Teacher and Base Student},
  colframe=caseBlue!65,
  colbacktitle=caseBlue!8
]
{\scriptsize Qwen3.6-35B / Base 30B Student\par}
\smallskip
{\footnotesize
\textbf{Success: 2/3 each}
\hfill Sample 1: Pass\par}

\begin{lstlisting}[
  style=repaircase,
  language={},
  escapeinside={(*@}{@*)}
]
@@ expand_record(rec)
    if f in rec:
        expanded_rec[f] = rec[f]

(*@\textcolor{resultPass}{\bfseries +   add\_db\_name(expanded\_rec)}@*)
    return expanded_rec
\end{lstlisting}

\tcblower
{\footnotesize
Connects the helper to record expansion,
so the required transformation executes.
\par}
\end{tcolorbox}
\end{minipage}
\hfill
\begin{minipage}[t]{0.49\textwidth}
\vspace{0pt}
\begin{tcolorbox}[
  repaircard,
  equal height group=openlibrarycomparison,
  title={SWE-Growth},
  colframe=caseGray!65,
  colbacktitle=caseGray!8
]
{\scriptsize Trained Qwen3-Coder-30B-A3B\par}
\smallskip
{\footnotesize
\textbf{Success: 0/3}
\hfill Sample 1: Fail\par}

\begin{lstlisting}[
  style=repaircase,
  language={},
  escapeinside={(*@}{@*)}
]
@@ utils/__init__.py
+def add_db_name(rec: dict) -> None:
+    ...  # constructs author identifier

(*@\textcolor{resultFail}{\bfseries \# expand\_record() was not changed.}@*)
\end{lstlisting}

\tcblower
{\footnotesize
Defines the helper but omits its call site,
leaving record-expansion behavior unchanged.
\par}
\end{tcolorbox}
\end{minipage}

\caption{
An Open Library retention failure
(\texttt{openlibrary-1351c59f}).
The Teacher and base Student each solve two of
three attempts; SWE-Growth solves none.
Excerpts show abbreviated sample-1 patches.
Green highlights the added helper call;
the red annotation identifies the missing integration
in SWE-Growth and is not part of the patch.
Ellipses denote omitted code.
}
\label{fig:openlibrary-retention-failure}
\end{figure*}

Figure~\ref{fig:openlibrary-retention-failure}
shows a task where SWE-Growth fails despite
successful repairs from both the frozen Teacher
and initial Student.
Their patches connect \texttt{add\_db\_name}
to \texttt{expand\_record}, ensuring that the
required transformation executes during record
expansion.
In the illustrated attempt, SWE-Growth defines
the helper but does not connect it to the calling
function, leaving the required behavior unchanged.
Across three attempts, SWE-Growth records no
successes, whereas both reference models succeed
twice.
This case illustrates that generating a relevant
helper is insufficient without integrating it
into the execution path, and that aggregate
improvements do not guarantee retention of
previously observed task-level successes.

\end{document}